\documentclass[12pt,prd]{revtex4}
\usepackage{graphicx}

\begin{document}

\title{Fundamental bound for time measurements and minimum uncertainty clocks}

\author{Rodolfo Gambini$^1$ and Jorge Pullin$^2$}
\affiliation{1. Instituto de F\'{\i}sica, Facultad de Ciencias, Igu\'a 4225, esq. Mataojo,
11400 Montevideo, Uruguay. \\
 2. Department of Physics and Astronomy, Louisiana State University,
Baton Rouge, LA 70803-4001.}

\begin{abstract}
We present a simple argument leading to a fundamental minimum 
uncertainty in the determination of times. It only relies in the
uncertainty principle and time dilation in a gravitational field. It
implies any attempt to measure times will have a fundamental level 
of uncertainty. Implications are briefly outlined. 
\end{abstract}
\maketitle

The issue of if the fundamental theories of physics impose limitations
on the accuracy of how we can determine physical quantities has been
analyzed by several authors over the years. Salecker and Wigner \cite{SaWi}
considered an idealized clock consisting of two mirrors with light
bouncing between them and concluded that the minimum uncertainty of
such a clock was proportional to the square root of the ratio of the
time to be measured and the mass of the clock. Such argument is based
exclusively on quantum mechanics. Several authors \cite{several} have
combined that argument with ingredients coming from gravity and
reached the conclusion that the clock's precision has a bound
proportional to fractional power of the time to be measured. In
particular the fact that one cannot concentrate arbitrarily large
quantities of energy in a finite region, as a black hole forms. The
constructions require the introduction of several elements, like
specific models for clocks or assumptions about the extent of regions
of strong gravitational fields. A separate argument by Ng and Lloyd \cite{nglloyd}
uses the Margolus--Levitin theorem and reaches similar conclusions. Here we would like to present a
streamlined argument that only relies on the uncertainty principle and
the time dilation in gravitational fields and basic error propagation
theory to put bounds on the precision of a clock.

The existence of fundamental limitations in the measurements of
physical quantities can have profound conceptual implications. For
instance, ordinary formulations of quantum mechanics treat time as a
classical variable, which implicitly implies that it can be measured
with arbitrary precision. Other variables are certainly not treated
this way. The limitations in time measurement may lead to a loss of
unitarity in the formulation, as variables measured by real clocks
cannot track the ideal classical time assumed in the formulation of
the Schr\"odinger equation \cite{pedagogical}.  In fact, limitations on the measurements
of space and time have led us to propose a new interpretation of
quantum mechanics, the Montevideo Interpretation \cite{review}.

We here consider a microscopic  quantum system playing the role of a
clock and a macroscopic  observer that interacts with the clock
interchanging signals. We start by considering the time-energy uncertainty relation,
\begin{equation}
  \label{eq:1}
  \Delta E \Delta t_c> \hbar.
\end{equation}
where $\Delta t_c$ is of the order of the period of oscillation of the
system being considered. The clock does not necessarily have to be
associated to a periodic motion. Busch {\em et al.} \cite{busch} have
proposed extensions that allow to consider time as an observable even
in the case of non periodic clocks. No matter what type of system we
consider, the Helstrom--Holevo bound \cite{helstrom} sets limits to
the measurement of the evolution time of any quantum state. The
uncertainty in any estimation of the evolution time of the state
through the measurement of an arbitrary observable satisfies the time
energy condition. 

We now consider the relationship between the time measured by the
clock locally, $t_c$, and an observer at an infinite distance from it,
$t$.  The gravitational time dilation was first described by Albert
Einstein in 1907 as a consequence of special relativity in accelerated
frames of reference. In general relativity, it is considered to be a
difference in the passage of proper time at different positions as
described by a the metric tensor of space-time. The relevance of this
effect in the determination of fundamental limitations to time
measurements was emphasized by Frenkel \cite{frenkel}. It is given by,
\begin{equation}
  \label{eq:2}
  t=\frac{t_c}{\sqrt{1-\frac{r_S}{r}}},
\end{equation}
with $r_S$ the Schwarzschild radius of the clock in question,
$r_s=2G E/c^4$ with $E$ the energy of the clock and $r$ its radius. We will assume that
an observer cannot be arbitrarily close to the clock. For a standard atomic clock this effect may seem
negligible but for an optimal clock it would be important as we shall
see. The best clocks we can consider are quantum systems that are
microscopic. In such a case an observer will not be able to get close
to the clock, and its distance, for all practical purposes can be
taken to be infinite (at least compared to the microscopic value of
$r$).

As an example, 
take an atomic clock based on the transition of an electron between
energy levels in an atom. We construct an electromagnetic source at
the
frequency that maximizes the probability that an electron transition
between levels. This is a frequency standard. The source emits photons
that interact with the atom and make the electron transition. We repeat
this for many atoms and check how many are excited. This allows to
maximize the transition probability. In reality the photon, after
being emitted, falls in the gravitational field of the atom, allowing
to use (2) to compute the red shift of the photon's frequency. There
is an uncertainty in the frequency of the emitted photon.
This  uncertainty will suffer the correction due to time dilation described above. 
Nowadays, the best clocks we have taken as reference are based
on low energy atomic transitions. It would be desirable to have
transitions of higher frequency, but stability presents an
experimental challenge. We are neglecting other possible quantum
gravitational effects since we are considering processes
that  always occur in clocks larger than their Schwarzschild radius, and we
will encode the fluctuations of the metric in the uncertainties of the
Schwarzschild radius.
\begin{figure}[h]
\includegraphics[width=11cm]{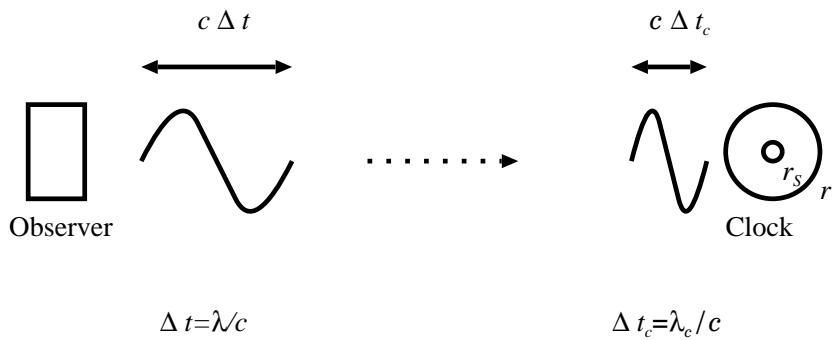} 
\caption{A schematic description of the setup}
\end{figure}

We would like to establish the uncertainty in the observed period of
oscillation.  Using the standard technique for the propagation of
errors of a measurement, taking differentials of the above expression, 
we have that,
\begin{equation}\label{4}
  \left(\Delta t\right)^2 = \frac{1}{4}\frac{t_c^2 \left(\Delta
      r_S\right)^2}{\left(1-\frac{r_S}{r}\right)^3 r^2}+\frac{\left(\Delta t_c\right)^2}{{1-\frac{r_S}{r}}},
\end{equation}
and from the definition of the Schwarzschild radius,
\begin{equation}
\left(\Delta r_S\right)^2=\frac{4 G^2 \left(\Delta E\right)^2}{c^8},
\end{equation}
and therefore for the clock that minimizes the uncertainty $\Delta t$
the following holds,
\begin{equation}
  \label{eq:3}
  \Delta r_S=\frac{4 G^2 \hbar^2}{c^8 \left(\Delta t_c\right)^2}.
\end{equation}
And substituting (\ref{eq:3}) in (\ref{4}), we get,
\begin{equation}
\left(\Delta t\right)^2=\frac{t_c^2 G^2
  \hbar^2}{\left(1-\frac{r_S}{r}\right)^3 r^2 c^8 \Delta
  t_c^2}+\frac{\left(\Delta t_c\right)^2}{1-\frac{r_S}{r}},
\end{equation}
this indicates that $\Delta t_c$ will be bounded since it appears in
the numerator and the denominator. We observe that $1-r_S/r$ is a
positive quantity less than one, since the size of the clock cannot be
smaller than its Schwarzschild radius, and using (\ref{eq:2}) to
translate $t_c$ to $t$, one has that,
\begin{equation}
\left(\Delta t\right)^2 = \frac{t_c^2 G^2 \hbar^2}{\left(1-\frac{r_S}{r}\right)^3 r^2 c^8
  \left(\Delta t_c\right)^2} + \frac{\left(\Delta
    t_c\right)^2}{1-\frac{r_S}{r}}> \frac{t^2 G^2 \hbar^2}{r^2 c^8
    \left(\Delta t_c\right)^2}+\left(\Delta t_c\right)^2,
\end{equation}
and assuming that the clock has size $r$ and that the oscillation
within it takes place at the mean
speed $v$ we have that  
$2\pi r=v \Delta t_c$. Then, 
differentiating with respect to $\Delta t_c$ to find the minimum
value of the time uncertainty and get, that the minimum occurs at,
\begin{equation}
\Delta t_c=\frac{\sqrt{2} \pi^{1/3} c^{1/3} \left(t^2 t_{\rm Planck}^4\right)^{1/6}}{v^{1/3}},
\end{equation}
with $l_{\rm Planck}$ Planck's length. This expression is minimized
when $v$ is the speed of light, yielding,
\begin{equation}
\Delta t_c=\sqrt{2} \pi^{1/3} \left(t t_{\rm Planck}^2\right)^{1/3},
\end{equation}
with $t_{\rm Planck}$ Planck's time, 
and this corresponds to a value of the error in the observed time of 
\begin{equation}
  \label{eq:4}
  \Delta t > \sqrt{3} \pi^{1/3} t^{1/3} t_{\rm Planck}^{2/3}.
\end{equation}
This bound
depends on $t$, for $t =1$ second is ten orders of magnitude smaller
than the accuracy of the best current clocks, $\Delta t/t$ is smaller
the larger the time $t$. One can also estimate $\Delta E/E$, which
turns out to be smaller than $10^{-20}$ for the measurement of a one
second interval and decreases as the interval to be measured
increases. Using equation  (2) it can be easily seen that this bound is saturated
by a clock with a radius approximately given by $r=3 r_S$.

This limit is similar to the ones obtained by previous authors but it
did not assume any particular model of the clock, only the uncertainty
principle and the formula for time dilation in a gravitational
field. It also suggests, taking into account equations (2), (9) and (10), what is the ideal clock, a clock that
saturates the bound, an oscillation given
by a particle orbiting the black hole near $3 r_S$, the innermost stable
circular orbit of a non-rotating black hole. It might be possible to
go beyond that with orbits at slightly lower radius, respecting the
bound
of equation (10), but it would be difficult
to create a stable system.

Summarizing, it is possible to obtain a bound for the accuracy
reachable by a clock using only fundamental bounds of quantum
mechanics and taking into account the gravitational time dilation
without having to go into the details of the practical implementation
of the clock. The analysis is interpretation independent.

We thank an anonymous referee for bringing the Helstrom--Holevo bound
argument to our attention.
This work was supported in part by Grants NSF-PHY-1603630,
NSF-PHY-1903799, funds of the Hearne Institute for
Theoretical Physics, CCT-LSU, Pedeciba and Fondo Clemente Estable
FCE\_1\_2019\_1\_155865.

\end{document}